\documentclass{article}

    \PassOptionsToPackage{numbers, compress}{natbib}

    \usepackage[preprint]{neurips_2022}

\usepackage[utf8]{inputenc} %
\usepackage[T1]{fontenc}    %
\usepackage{hyperref}       %
\usepackage{url}            %
\usepackage{booktabs}       %
\usepackage{amsfonts}       %
\usepackage{nicefrac}       %
\usepackage{microtype}      %
\usepackage{xcolor}         %
\usepackage{subcaption}
\usepackage{lipsum}
\usepackage{dsfont}
\usepackage{graphicx}
\usepackage{float}
\usepackage{amsmath}
\usepackage{amsthm}
\usepackage{enumitem}
\usepackage{algorithm}
\usepackage{xspace}
\usepackage{multirow}
\usepackage[export]{adjustbox}
\usepackage[framemethod=TikZ]{mdframed}
\usepackage[noend]{algpseudocode}
\usepackage[strings]{underscore}

\newtheorem{theorem}{Theorem}[section]
\newtheorem{lemma}[theorem]{Lemma}

\newtheorem{definition}[theorem]{Definition}

\newtheorem{fact}[theorem]{Fact}

\renewenvironment{quote}
  {\list{}{\rightmargin=0.4cm \leftmargin=0.4cm}%
   \item\relax}
  {\endlist}

\NewDocumentCommand\p{ m g }{
  \ensuremath{
    \IfNoValueTF{#2}
    {\Pr [ #1 ]}
    {\Pr_{#1}[#2]}
  }
}
\RenewDocumentCommand\P{ m g }{
  \ensuremath{
    \IfNoValueTF{#2}
    {\Pr \left[#1\right]}
    {\Pr_{#1}\left[#2\right]}
  }
}
\DeclareMathOperator*{\Expectation}{\mathbb{E}}
\NewDocumentCommand\E{ m g }{
  \ensuremath{
    \IfNoValueTF{#2}
    {\Expectation \left[#1\right]}
    {\Expectation_{#1}\left[#2\right]}
  }
}
\DeclareMathOperator*{\Variance}{\mathbb{V}\mathrm{ar}}
\NewDocumentCommand\Var{ m g }{
  \ensuremath{
    \IfNoValueTF{#2}
    {\Variance \left[#1\right]}
    {\Variance_{#1}\left[#2\right]}
  }
}

\newcommand{\UsrData}[1]{x_{#1}}
\newcommand{\UsrRndData}[1]{y_{#1}}
\newcommand{\UsrSet}[1]{S_{#1}}

\NewDocumentCommand\dyadicInterval{ g g }{
  \ensuremath{
    \IfNoValueTF{#2}
    { \IfNoValueTF{#1}{ \mathcal{I} }{ \mathcal{I}_{#1, * } } }
    { \mathcal{I}_{#1, #2} }
  }
}
\NewDocumentCommand\dyadicIntervalSet{g}{
  \ensuremath{
    \IfNoValueTF{#1}{ \textit{ISet} }{ \textit{ISet}[{#1}] }
  }
}

\NewDocumentCommand\partialsum{gg}{
  \ensuremath{
    \IfNoValueTF{#2}
    { S ({#1} ) }
    { S_{#1} ({#2}) }
  }
}

\newcommand{\IntSet}[2]{[{#1}\,.\,.\,{#2}]}

\newcommand{\norm}[1]{\left\Vert {#1} \right\Vert}
\newcommand{\PAREN}[1]{{\left( {#1} \right)}}
\newcommand{\paren}[1]{{( {#1} )}}
\newcommand{\angles}[1]{{\langle {#1} \rangle}}
\newcommand{\bracket}[1]{{\left[ {#1} \right]}}

\newcommand{\card}[1]{\left| {#1} \right|}
\newcommand{\set}[1]{\left\{ {#1} \right\}}
\newcommand{\eps}{\varepsilon}

\newcommand{\basis}[1]{\vec{e}_{#1}}

\newcommand{\Lb}{\text{LB}}
\newcommand{\Ub}{\mathsf{UB}}

\def\algoPerfectPartitioning{\mathcal{A}_{ \textit{infinite} }}
\def\algoFiniteTime{ \mathcal{A}_{ \textit{rnd-wlk} } }
\newcommand{\algoSampleThenWalk}{\mathbf{A}_{ \textit{smpl-wlk} }}

\newcommand{\cA}{\mathcal{A}}

\newcommand{\cC}{\mathcal{C}}
\newcommand{\cD}{\mathcal{D}}

\newcommand{\cR}{\mathcal{R}}

\newcommand{\cX}{\mathcal{X}}
\newcommand{\cY}{\mathcal{Y}}
\newcommand{\cZ}{\mathcal{Z}}

\newcommand{\N}{\mathbb{N}}
\newcommand{\R}{\mathbb{R}}

\newcommand{\mbfA}{\mathbf{A}}

\newcommand{\mbfD}{\mathbf{D}}
\newcommand{\mbfP}{\mathbf{P}}

\newcommand{\sDist}{\vec{\pi}}
\NewDocumentCommand\vecp{g}{
  \ensuremath{
    \IfNoValueTF{#1}{ \vec{p} }{ \vec{p}^{\,(#1)} }
  }
}

\newif\ifcomment
\commentfalse

\ifcomment
\newcommand{\tony}[1]{\textcolor{brown}{[TONY: #1]}}
\newcommand{\hao}[1]{\textcolor{blue}{[HAO: #1]}}
\newcommand{\olya}[1]{\textcolor{purple}{[Olya: #1]}}

\definecolor{DarkGreen}{rgb}{0.1,0.5,0.1}

\definecolor{DarkGreen}{rgb}{0.1,0.5,0.1}

\else
\newcommand{\tony}[1]{}
\newcommand{\hao}[1]{}
\newcommand{\olya}[1]{}
\fi

\title{Walking to Hide: Privacy Amplification via Random Message Exchanges in Network}

\author{%
    Hao Wu \qquad Olga Ohrimenko \qquad Anthony Wirth \\
    School of Computing and Information Systems \\
    The University of Melbourne \\
    \texttt{whw4@student.unimelb.edu.au}\\
    \texttt{\{oohrimenko,awirth\}@unimelb.edu.au}\\
}

\begin{document}

\maketitle

\begin{abstract}
    The \emph{shuffle model}
    is
    a powerful tool to amplify the privacy guarantees of the \emph{local model} of differential privacy.
    In contrast to the fully decentralized manner of guaranteeing privacy in the local model, the shuffle model requires a central, trusted shuffler.
    To avoid this central shuffler, recent work of \citeauthor{Liew2022}~(2022) proposes shuffling locally randomized data in a decentralized manner, via random walks on the communication network constituted by the clients. The privacy amplification bound it thus provides depends on the topology of the underlying communication network, even for infinitely long random walks.
    It does not match the state-of-the-art privacy amplification bound for the shuffle model (\citeauthor{FeldmanMT21},~\citeyear{FeldmanMT21}).
    
    In this work, we prove that the output of~$n$ clients' data, each perturbed by a~$\eps_0$-local randomizer, and shuffled by random walks
    with a logarithmic number of steps, is~$\paren{ {O}\paren{ \paren{1 - e^{-\eps_0}}\sqrt{\paren{e^{\eps_0} / n} \ln \paren{1 / \delta }}}, O(\delta)}$-differentially private.
    Importantly, this bound is independent of the topology of the communication network, and  asymptotically closes the gap between the privacy amplification bounds for the network shuffle model (\citeauthor{Liew2022}, 2022) and the shuffle model (\citeauthor{FeldmanMT21},~\citeyear{FeldmanMT21}). 
    Our proof is based on a reduction to the shuffle model, and an analysis of the distribution of random walks of finite length. 
    Building on this, we further show that if each client is sampled independently with probability~$p$, the privacy guarantee of the network shuffle model can be further improved to~$\paren{ {O}\paren{ \paren{1 - e^{-\eps_0}}\sqrt{p \paren{e^{\eps_0} / n} \ln \paren{1 / \delta }}}, O(\delta)}$.
    Importantly, the subsampling is also performed in a fully decentralized manner that  does not require a trusted central entity; compared with related bounds in prior work, our bound is stronger.
\end{abstract}
\section{Introduction}
\label{sec: introduction}

With the collection of data,  expanding in scale and variety, tech companies develop new products and improve existing services. This, however, poses unprecedented privacy risk for sensitive personal information. 
With increasing public scrutiny and consciousness of privacy protection, governments  enforcing stringent regulations~\citep{voigt2017eu}, considerable and consistent research effort has designed private data collection and analysis protocols.

Differential privacy~\citep{DworkMNS06, Dwork06} is emerging as the de-facto standard. 
A protocol is said to be \emph{differentially private} if its output varies only a little with individual data changes, and therefore little can be inferred about specific personal information. 
On the other hand, the protocol can still work with aggregated population information, admitting analysis of overall statistics.

Such protocols inherently add noise components to the released data.
Early studies focus on two models for releasing the noisy outputs, \emph{central}~\citep{DworkMNS06} and \emph{local}~\citep{KasiviswanathanLNRS11}. 
The central model assumes a trusted curator that collects the users' raw data, performs the aggregation, and publishes the (noisy) result.
In contrast, the local model is fully decentralized, and requires only a minimal trust assumption.
Each user manages their own data, and reports only a randomized version to the un-trusted curator.
Since the users inject a \emph{reasonable} amount of noise into their reports, the local model provides a stronger privacy guarantee, but much weaker data accuracy. 

\paragraph*{Shuffle models}
To mitigate the limitation of the local model, there has been a line of research on the \emph{shuffle} model~\citep{EFMRTT19, CheuSUZZ19, BalleBGN19, BalleKMTT20, FeldmanMT21}. 
As in the local model, the users report only a randomized version of their data.
These data are sent to a secure intermediary, a \emph{shuffler}, before being reported to the curator.
The shuffler performs a random permutation over the data to de-link them from users. 
The additional anonymity of the shuffler amplifies the privacy guarantee of the protocol, allowing the users to inject less noise into their local reports.  There is ongoing research into obtaining a tight amplification bound for the shuffle model~\citep{EFMRTT19, CheuSUZZ19, BalleBGN19, BalleKMTT20},  culminating in the work of \citeauthor{FeldmanMT21}~\cite{FeldmanMT21}. They state that  if the local algorithm~\footnote{One needs to assume that users apply the same local algorithm if shuffling is applied to the randomized data.} (called a local randomizer) with which~$n$ users randomize their data is $\eps_0$-differentially private (defined formally in Section~\ref{sec: problem definition}), the protocol that outputs the shuffled reports is~$\paren{O\paren{ \paren{1 - e^{-\eps_0}}\sqrt{\paren{e^{\eps_0} / n} \cdot \ln \paren{1 / \delta}}}, \delta}$-differentially private.

Crucially, the shuffle model relies on a \emph{trusted} central shuffler, impending the benefit of achieving the privacy guarantee in a fully decentralized manner, as in the local model.
Alternative approaches~\citep{CB22, Liew2022} avoid reintroduction of such central shuffler, and try to achieve the privacy amplification in a fully decentralized manner.
These approaches~\citep{CB22, Liew2022} rely on peer-to-peer exchanges of messages between users, modelling the communication network as a graph where the users and the communication channels between them are viewed as the vertices and the edges, respectively.
Computations are performed by walks on such network.
\citeauthor{CB22}~\citep{CB22} pioneered this approach, introducing a relaxed definition of local differential privacy. They showed the corresponding privacy guarantees of some data queries can be improved when the computation can be modelled by a deterministic walk on a ring or a random walk on a complete graph. Recent work by~\citeauthor{Liew2022}~\citep{Liew2022} showed that the privacy guarantee can be amplified if the perturbed data generated by the users are exchanged in a random-walk-driven manner on the graph. 
Their result does not depend on the relaxed definition of local differential privacy, imposes no restriction on the query type, and little restriction on the graph type. 
If the local randomizer is $\eps_0$-differentially private, then their model achieves a privacy guarantee of 
\begin{equation}
    \paren{ O\paren{ (\eps_1 + \eps_2) \cdot \paren{e^{\eps_0} - 1} e^{2\eps_0} \sqrt{\ln \paren{1 / \delta}} }, \delta + \delta' }\,, 
\end{equation}
where $\eps_1 = O \paren{ \sqrt{\sum_{i \in [n]} \paren{\sDist[i]}^2+ (1 - \alpha)^{2T} } }$, $\eps_2 \in O \paren{  \sqrt{\paren{1 / n} \cdot \ln \paren{1  / \delta'}}}$, and where~$\sDist \in \R^n$ is the stable distribution of the random walks, which depends on the topology of the graph, and where~$\alpha \in (0, 1)$ is the spectral gap, and~$T \in \N^+$ is the number of steps of the random walks. 
The work~\citep{Liew2022} also explores subsampling to further enhance the privacy guarantee to~$\paren{O\paren{ e^{\eps_0} \paren{e^{\eps_0} - 1} \eps_1 \sqrt{  \ln \paren{1 / \delta} } }, \delta}$. 
Crucially, such bounds depend on~$\sDist$.
It can be proven that there exists a graph such that~$\sum_{i \in [n]} \paren{\sDist[i]}^2 \in \Theta \paren{1}$, in which case the bounds provide no amplification at all. 
Even in the best case, it holds that~$\sum_{i \in [n]} \paren{\sDist[i]}^2 \ge1 / n$, such bounds are still inferior to the best bound,~$\paren{ \tilde{O}\paren{ \paren{1 - e^{-\eps_0}}\sqrt{\paren{e^{\eps_0} / n} }}, \delta}$, obtained under shuffle model of~\citeauthor{FeldmanMT21}~\citep{FeldmanMT21}.

\begin{quote}
    \textbf{Research Question:}
    \emph{Can we obtain amplification bounds for the network shuffle model that do not depend on $\sDist$, and close the gap of the amplifications bounds between the \emph{network shuffle} model and the \emph{shuffle} model?}
\end{quote}

\noindent {\bf Our Contributions.}
In this work, we answer the research question in the affirmative. 
We prove that the network shuffle model, for a suitable choice of~$T$, achieves a privacy guarantee of~$\paren{ \tilde{O}\paren{ \paren{1 - e^{-\eps_0}}\sqrt{\paren{e^{\eps_0} / n} }}, \delta}$, which is independent of~$\sDist$, and asymptotically the same as the state-of-the-art bound in the shuffle model. 
Our proof provides a new perspective of network shuffle, by viewing it as a \emph{decentralized implementation} of the shuffler. 
We start by showing that if the number of steps of the random walk,~$T$, approaches infinity, the output distribution of the network shuffle model converges to the distribution of a function which is a post-processing of the central shuffler. Therefore, perfect shuffling is achieved and the privacy amplification bound by~\citeauthor{FeldmanMT21}~\citep{FeldmanMT21} applies directly for this ideal scenario. 
We then show that for a finite, but suitable, choice of~$T$, the output distribution of the network shuffle model does not deviate significantly  from the one when~$T \rightarrow \infty$, and the privacy amplification bound remains the same.

We further present a new protocol,
which combines Poisson sampling and the network shuffling model.
We prove that when each client reports their perturbed datum independently with probability~$p$, the network shuffle achieves privacy guarantee of~$\paren{ \tilde{O}\paren{ \paren{1 - e^{-\eps_0}} \sqrt{ p \cdot e^{\eps_0} / n}}, \delta}$. 
Since the network shuffle model is a decentralized implementation of the central shuffler, such bound also applies to the \emph{shuffle model} when combined with subsampling. 

\noindent {\bf Organization.}
Our paper is organized as follows.
Section~\ref{sec: problem definition} introduces the model, formally. 
Section~\ref{sec: preliminaries} discusses the preliminaries for our proof.
Section~\ref{sec: proof of privacy amplification} develops the privacy guarantee for the network shuffle model. 
Section~\ref{sec: subsampling} shows how subsampling can further improve the privacy guarantee.
Section~\ref{sec: review} summarizes the related works.
\section{Model Description}
\label{sec: problem definition}

In this model, there is a untrusted curator and a set of~$n$ users. 
Each user $u \in [n]$ holds a datum,~$x_u$, from some domain~$\cX$. 
A dataset~$x_{1:n} \doteq \paren{x_1, \ldots, x_n} \in \cX^n$ is an ordered collection of the~$n$ elements. 

{\bf Differential Privacy.}
Let~$x_{1:n}' \doteq \paren{x_1', \ldots, x_n'} \in \cX^n$ be another possible dataset.

\begin{definition}[Neighboring Datasets]
    Two datasets~$x_{1:n}, x_{1:n}' \in \cX^n$ are called \emph{neighboring}, denoted by~$x_{1:n} \sim x_{1:n}'$, if~$\Vert x_{1:n} - x_{1:n}' \Vert_0 \le 1$, i.e., they differ for at most one user. 
\end{definition}

In order for a randomized algorithm to hide %
sensitive information of an individual datum, its output distributions should be similar on neighboring inputs.
Formally, we require it to be~$\paren{\eps, \delta}$-\emph{differentially private}, which is famously defined as follows.

\begin{definition}[$\paren{\eps, \delta}$-Private Algorithm~\citep{DR14}] \label{def: Differential Privacy}
    Given~$\eps, \delta > 0$, a randomized algorithm $\cA: \cX^n \rightarrow \cZ$ is called~$\paren{\eps, \delta}$-differentially private, if for all neighboring datasets~$x_{1:n}, x_{1:n}' \in \cX^n$ and all (measurable) $Z \subseteq \cZ$,
    \begin{equation} \label{ineq: def private algo}
        \P{ \cA (x_{1:n}) \in Z } \le e^\eps \cdot \Pr [ \cA (x_{1:n}') \in Z ] + \delta\,.
    \end{equation}
\end{definition}

We call~$\eps$ and~$\delta$ the \emph{privacy parameters}.
Typically, it is required that~$\delta \ll 1 / n$~\citep{DR14}. 
The model also involves local algorithms, which we call \emph{local randomizers}, with which each user perturbs their own datum. 
In a similar vein, we can define~$\paren{\eps, \delta}$-differential privacy for local randomizers, viz.

\begin{definition}[$\paren{\eps, \delta}$-Local Randomizer~\citep{DR14}] \label{def: loccal randomizer}
    Given~$\eps, \delta > 0$, a local randomizer~$\cR: \cX \rightarrow \cY$ is~$\paren{\eps, \delta}$-differentially private, if for all~$x, x' \in \cX$, and all (measurable) $Y \subseteq \cY$, 
    \begin{equation} \label{ineq: def local randomizer}
        \P{ \cR (x) \in Y } \le e^\eps \cdot \Pr [ \cR (x') \in Y ] + \delta\,.
    \end{equation}
\end{definition}

{\bf Network Shuffling Model.}
The model proposed by~\citeauthor{Liew2022}~\citep{Liew2022} is motivated by the fact that in many applications, the clients can communicate not only with the curator, but also with some other clients using a secure channel (e.g., using end-to-end encryption). 
The network created from such client-to-client communication channels can be modelled by a graph~$G = (V, E)$, where the clients are viewed as the vertices, and the communication channels between them as the edges.

{\it Model Assumptions:} the model makes the following assumptions: 1) no collusion between the clients and the server 2) clients are honest but curious, i.e., the clients execute the protocol faithfully; 3) the adversary may know the graph topology, but it cannot track the traffic of the clients' communication.

The network shuffling protocol, denoted as~$\algoFiniteTime$, works as follows:

\begin{figure}[!t]
	\centering
	\includegraphics[width=\linewidth]{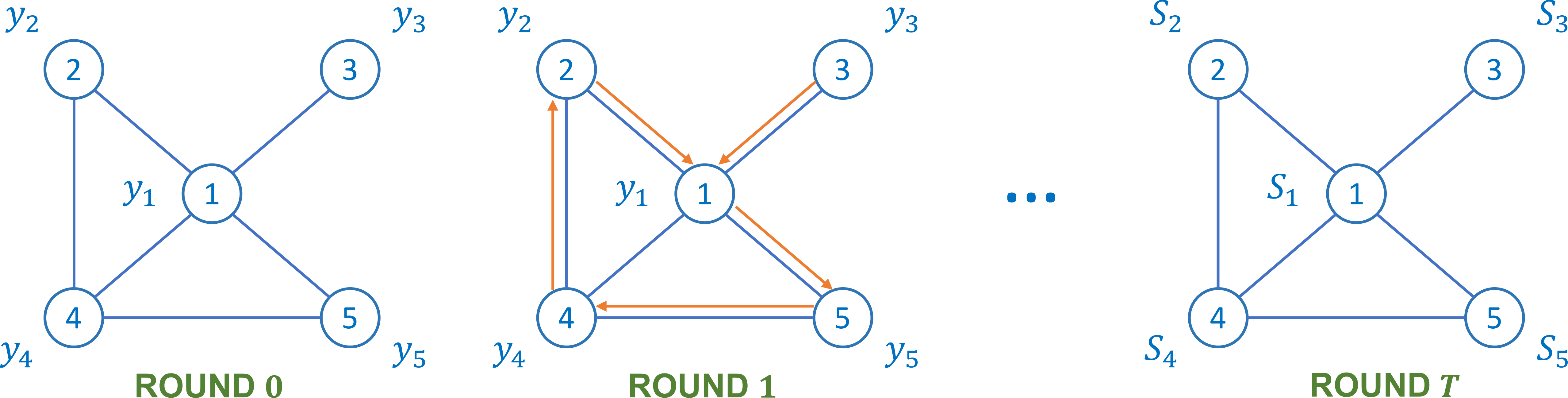}
	\vspace{0.5mm}
	\caption{
	    A running example of \emph{network shuffling model}. Initially, at round~$0$, the five clients generate their perturbed data~$y_u, u \in [5]$. 
	    In round~$1$, the~$y_u$ start their random walks, each arrow indicates the direction of the first step. 
	    The random walks continue for~$T$ rounds. %
	    At the end of round~$T$, for each~$u \in [n]$, $S_u$ is the multi-set of perturbed data received by client~$u$. 
	}
	\label{fig: random walk}
\end{figure}

\begin{enumerate}
    \item Initially, each client~$u \in [n]$ perturbs its datum~$\UsrData{u}$ with an~$\paren{\eps_0, \delta_0}$-local randomizer~$\cR : \cX \rightarrow \cY$, whose output is denoted by~$\UsrRndData{u} \doteq \cR(\UsrData{u})$.
    Note that different clients invoke the same local algorithm, but with independent randomness. 
    
    \item The perturbed datum~$\UsrRndData{u}$ starts a~$T$-step random walk from client~$u$.
    
    \item When the random walks of all perturbed data stop, each client~$u \in [n]$ reports to the curator the multiset~$\UsrSet{u} \in \paren{\N}^{\cY}$ of data they received, where~$\paren{\N}^{\cY}$ comprises all possible multisets from~$\cY$. 
\end{enumerate}

An example of~$\algoFiniteTime$ is shown in Figure~\ref{fig: random walk}, and the client-side protocol for~$\algoFiniteTime$ is described in Algorithm~\ref{algo: finite time random walk}.
Note that the output of~$\algoFiniteTime$ (which is revealed to the curator) is a collection of multisets~$\set{\UsrSet{u}, u \in [n]}$, which constitutes a partition of the perturbed data~$\set{ \UsrRndData{u} : u \in [n]}$.
There are multiple feasible representations for each multiset~$\UsrSet{u}$:
if the range~$\cY$ of local randomizer~$\cR$ is finite, $\UsrSet{u}$ can be represented by a histogram over~$\cY$; 
or~$\UsrSet{u}$ can be represented by just a random permutation of elements it contains.
Importantly~$\UsrSet{u}$ should be a pure multiset, containing no information about the order in which~$u$ received information in the last round.

Our aim is to analyze the privacy guarantees of the outputs~$\set{\UsrSet{u}, u \in [n]}$ of~$\algoFiniteTime$. 
In broad terms, the sources of privacy protection by~$\algoFiniteTime$ are threefold: the noise components injected by the local randomizers; the data anonymization provided by random walks; and the further anonymization of each client returning the data they received as a pure multiset.

\begin{algorithm}[!ht]
    \caption{Client-Side Protocol for~$\algoFiniteTime$}
    \label{algo: finite time random walk}
    \begin{algorithmic}[1]
        \Require User Datum~$\UsrData{u} \in \cD$; Round Bound~$T \in \N^+$.
        \State $\UsrRndData{u} \leftarrow \cR\paren{\UsrData{u}}$.
        \State $\UsrSet{u} \leftarrow \{ \UsrRndData{u} \}$ %
        \label{line: algo: finite time random walk: set report set}
        \For{$i \in [T]$} 
        \label{line: algo: finite time random walk: walk start}
            \While{$\UsrSet{u} \neq \emptyset$} 
                \State Let~$y$ be an arbitrary element in~$\UsrSet{u}$
                \State Send~$y$ to one of~$u$'s neighbors uniformly at random. 
                \State Remove~$y$ from~$\UsrSet{u}$ \Comment{Remove just one instance of~$y$} %
            \EndWhile
            \State Receive reports from neighbors, and add them to~$\UsrSet{u}$. %
        \EndFor
        \State \Return $\UsrSet{u}$ to the curator as a pure multiset.
        \label{line: algo: finite time random walk: algo end}
    \end{algorithmic}
\end{algorithm}

\section{Preliminaries}
\label{sec: preliminaries}

In this section, we present additional preliminaries required for our privacy analysis. 

\vspace{-2mm}
\subsection{Post-Processing}

Post-processing the output of a differentially private algorithm preserves its privacy guarantee, viz.

\begin{fact}[~\citep{DR14}]
    \label{fact: post processing}
    Let~$\cA: \cX^n \rightarrow \cZ$ be a randomized algorithm that is~$\paren{\eps,\delta}$-differentially private.
    Let $f : \cZ \rightarrow \cZ'$ be an arbitrary randomized mapping. 
    Then $f \circ \cA: \cX^n \rightarrow \cZ'$ is~$\paren{\eps,\delta}$-differentially private.
\end{fact}

\subsection{Random Walk}

We consider a \emph{random walk} on an undirected and connected graph $G = \angles{V, E}$, with $n = |V|$ vertices and $m = |E|$ edges.
For convenience, number the vertices arbitrarily from~$1$ to~$n$. 
The random walk starts from a vertex, which is chosen deterministically, or according to some probability distribution over~$V$.
At each step, it picks uniformly at random one of the edges incident on the current vertex, and moves to the other endpoint of that edge.
For each $t \in \N$, let~$\vecp_t \in \R^n$ be the \emph{position probability distribution}, such that~$\vecp_t[u]$ is the probability that walk is at vertex~$u$ at the $t^{\text{th}}$ step (and therefore~$\sum_{u \in [n]} \vecp_t[u] = 1$).
In particular, $\vecp_0$ is the initial distribution. 
In the case the walk starts deterministically at a vertex~$u \in [n]$, we have~$\vecp_0 = \basis{u}$, where~$\basis{u}$ is the~$u^{\text{th}}$ standard basis vector.

{\bf Adjacency Matrix.}
Define the \emph{adjacency matrix} of~$G$ to be $\mbfA \in \set{0, 1}^{n \times n}$, such that, for each $u, v \in [n]$,~$\mbfA_{u, v}$ equals~$1$ if~$\paren{u, v} \in E$ and~$0$ otherwise. Since~$G$ is undirected, the matrix~$\mbfA$ is symmetric. 
The \emph{degree matrix}~$\mbfD \in \R^{n \times n}$ is a diagonal matrix, such that for each $u \in [n]$, $\mbfD_{u, u} = d_u$, where the degree~$d_u$ is the number of edges incident on vertex~$u$. 
Finally, define the \emph{transition matrix}~$\mbfP \in [0, 1]^{n \times n}$ by~$\mbfP \doteq \mbfA \mbfD^{-1}$. 
Since~$\mbfP$ is also symmetric, it has real eigenvalues~$\alpha_1 \ge \alpha_2 \ge \cdots \ge \alpha_n$, counted with multiplicities and sorted in non-increasing order. 

\begin{definition}[Spectral Gap~\citep{Williamson2016}]
    \label{def: spectral gap}
    $\alpha \doteq \min \set{1 - \alpha_2, 1 - \card{\alpha_n}}$ is known as the \emph{spectral gap} of the graph~$G$. 
    When~$G$ is connected and non-bipartite, it is guaranteed that~$\alpha < 1$.    
\end{definition}

Now, for each~$t \in \N$, $\vecp_t$ can be expressed concisely as~$\vecp_t =  \mbfP^t \vecp_0$. 
We are interested in the long-term behaviour of~$\vecp_t$.

\begin{definition}[Stationary Distribution]
    A probability distribution $\sDist$ is a \emph{stationary distribution} for the graph~$G$  if $\sDist = \mbfP \sDist$.
\end{definition}
    The distribution~$\sDist = \paren{\frac{d_1}{2m}, \frac{d_2}{2m}, \ldots, \frac{d_n}{2m}}$ is clearly stationary.
Under some conditions, for each initial distribution~$\vecp_0$, the distribution~$\vecp_t$ converges pointwise
to a unique stationary distribution. 
In such case,~$\paren{\frac{d_1}{2m}, \frac{d_2}{2m}, \ldots, \frac{d_n}{2m}}$ is the unique stationary distribution (for undirected graph).  

\begin{definition}[Ergodic Random Walk~\citep{Williamson2016}]
    A random walk is ergodic if there exists a stationary distribution~$\sDist$ such that for each initial distribution~$\vecp_0$, we have~$\lim_{t \rightarrow \infty} \vecp_t = \sDist$.
\end{definition}

The condition for which a random walk is ergodic, and the convergence rate of~$\vecp_t$ is stated as follows. 

\begin{fact}[Convergence~\citep{Williamson2016}]
    \label{fact: mixing time}
    A random walk on an undirected graph~$G$ is ergodic if and only if~$G$ is connected and non-bipartite.
    Further, 
    \begin{equation}
        \norm{\vecp_t - \sDist}_1 \le \sqrt{n} \paren{1 - \alpha}^t\,.
    \end{equation}
\end{fact}

\section{Privacy Amplification of Network Shuffle}
\label{sec: proof of privacy amplification}

In this section, we present our
 improved analysis for the privacy amplification bound of~$\algoFiniteTime$.
The main result of our paper is summarized as follows:

\begin{theorem}[Privacy Amplification by~$\algoFiniteTime$]
    \label{theorem: privacy amplification by finite random walk}
    Let $\cR : \cX \rightarrow \cY$ be an arbitrary $\paren{\eps_0, \delta_0}$-differentially private local randomizer.
    Suppose~$\algoFiniteTime$ runs the random walks for~$T = \paren{1 / \alpha} \ln \paren{n^{4.5} / \eps_0}$ rounds, where~$\alpha$ is the spectral gap of~$G$ (Definition~\ref{def: spectral gap}). 
    Then for every~$\delta \in [0, 1]$ such that~$\eps_0 \le \log \paren{ \frac{n}{16 \log \paren{2 / \delta}} }$,~$\algoFiniteTime$ is~$\paren{\eps, e^{\eps_0 / \paren{2n}} \delta'}$-differentially private, where
    \begin{equation}
        \label{equa: def of delta'}
        \delta' \doteq \delta + \paren{e^\eps + 1}\paren{1 + e^{-\eps_0} / 2} n \delta_0   
    \end{equation}
    and 
    \begin{equation}
        \label{ineq: privacy amplificaiton via finite time random walk}
        \eps \le \frac{\eps_0 }{n} + \log 
        \PAREN{1 + \frac{e^{\eps_0} - 1}{e^{\eps_0} + 1}                  \PAREN{
                \frac{
                    8 \sqrt{e^{\eps_0} \log \paren{4 / \delta}}
                }{
                    \sqrt{n}
                } + 
                \frac{
                    8 e^{\eps_0}
                }{
                    n
                }
            }
        }
        .
    \end{equation}
    In particular, when~$\cR$ is~$\eps_0$-differentially private, then $\algoFiniteTime$ is~$(\eps, e^{\eps_0 / \paren{2n}} \delta)$-differentially private. 
\end{theorem}

Our proof of Theorem~\ref{theorem: privacy amplification by finite random walk} comprises two steps. 
First, in Section~\ref{subsec: infinite step random walks}, we prove that if the random walk length~$T$ approaches infinity, the output distribution of the network shuffle model converges to the output distribution of a function, which can be viewed a post-processing of the shuffle model. 
According to Fact~\ref{fact: post processing}, whatever privacy guarantee is achieved by the shuffle model is achieved by this function. 
It follows that the state-of-the-art privacy amplification bound for the shuffle model by~\citeauthor{FeldmanMT21}~\citep{FeldmanMT21} applies directly for this function, and therefore for the network shuffle model. 

Next, in Section~\ref{subsec: finite step random walk}, we show that for finite but suitable choice of~$T$, the output distribution of the network shuffle model does not deviate significantly from the one where~$T =\infty$, and that the privacy guarantee remains the same.

\subsection{Infinite-Step Random Walk}
\label{subsec: infinite step random walks}

{\bf Basic Building Block.} 
Our proof relies on a fundamental privacy amplification result by~\citeauthor{FeldmanMT21}~\citep{FeldmanMT21}. 
In their model, a random permutation is performed over the raw data before the local randomizers are applied to them. This contrasts with the commonly studied \emph{shuffle} model~\citep{CheuSUZZ19, BalleBGN19}, where the random permutation is performed after the local randomizers are applied to the raw data. 
This enables~\citeauthor{FeldmanMT21}~\citep{FeldmanMT21} to analyze the case where the local randomizers are chosen adaptively: the output of a local randomizer depends not only on its current input, but possibly the previous outputs of the local randomizers.

\begin{fact}[Privacy Amplification by Shuffling~\citep{FeldmanMT21}]
    \label{fact: privacy amplification via uniform shuffling}
    For every domain~$\cX$, let $\cR^{(u)} : \cY^{(1)} \times \cdots \times \cY^{(u-1)} \times \cX \rightarrow \cY^{(u)}$ for~$u \in [n]$, where~$\cY^{(u)}$ is the range of~$\cR^{(u)}$, be a sequence of algorithms such that~$\cR^{(u)}(y_{1:u-1}, \cdot)$ is an~$\paren{\eps_0, \delta_0}$-local randomizer for all values of auxiliary inputs~$y_{1:u-1} \doteq \paren{y_1, \ldots, y_{u- 1} } \in \cY^{(1)} \times \ldots \times \cY^{(u-1)}$.
    Let $\cA' : \cX^n \rightarrow \cY^{(1)} \times \ldots \times \cY^{(n)}$ be the algorithm that, given a dataset $\paren{x_1, \ldots, x_n} \in \cX^n$, samples a uniform random permutation~$\sigma$ over~$[n]$, then sequentially computes~$y_u \doteq \cR^{(u)}(y_{1:u-1}, x_{\sigma(u)})$ for~$u \in [n]$ and outputs~$y_{1:n}$.
    For every~$\delta \in [0, 1]$ such that~$\eps_0 \le \log \paren{ \frac{n}{16 \log \paren{2 / \delta}} }$,~$\cA'$ is~$\paren{\eps, \delta'}$-DP, where~$\delta'$ is as in Equation~(\ref{equa: def of delta'}) and 
    \begin{equation}
        \label{ineq: privacy amplificaiton via uniform shuffling}
        \eps \le \log 
        \PAREN{1 + \frac{e^{\eps_0} - 1}{e^{\eps_0} + 1}                  \PAREN{
                \frac{
                    8 \sqrt{e^{\eps_0} \log \paren{4 / \delta}}
                }{
                    \sqrt{n}
                } + 
                \frac{
                    8 e^{\eps_0}
                }{
                    n
                }
            } 
        }\,.
    \end{equation}
\end{fact}

Observe that, if we replace each local randomizer~$\cR^{(u)}, u \in [n]$ by a common randomizer~$\cR: \cX \rightarrow \cY$ (with independent randomness for different~$u \in [n]$),
 shuffling of $\paren{\cR(x_1), \ldots, \cR(x_n)}$ is distributed the same as $\paren{\cR(x_{\sigma(1)}), \ldots, \cR(x_{\sigma(n)})}$, where~$\sigma$ is a uniform random permutation of~$[n]$. 
Therefore the amplification bound stated in Fact~\ref{fact: privacy amplification via uniform shuffling}
  still applies. 

{\bf Random Walk as Partitioning.}
We continue to show, with infinitely many rounds of random walk, we can reduce the analysis for the network shuffling model  to the analysis for the standard shuffle model. 
Consider the perturbed datum,~$\UsrRndData{u}$, generated by client~$u \in [n]$. 
The probability distribution of its position after a~$t$-step random walk can be computed by~$\mbfP^t \basis{u}$, where~$\basis{u}$ is the~$u^{(th)}$ standard basis vector. 
Suppose that the undirected graph~$G$ is connected and non-bipartite.
According to Fact~\ref{fact: mixing time}, %
~$\lim_{t \xrightarrow{} \infty} \mbfP^t \basis{u} = \sDist$.
Therefore, \emph{the net effect of an infinite-step random walk is that the datum~$\UsrRndData{u}$ appears in a node~$v$ with probability~$\sDist[v]$, for each~$v \in [n]$}.
Further, for different~$u$, the positions of the~$\UsrRndData{u}$ are independent. 
This motivates our analyzing
the following idealized model, the pseudo-code for which is given in Algorithm~\ref{algo: inifinite time random walk}.
\begin{enumerate}
    \item Initially, each client~$u \in [n]$ perturbs its datum,~$\UsrData{u}$, with a local randomizer~$\cR$, whose output is denoted by~$\UsrRndData{u} \doteq \cR(\UsrData{u})$.
    
    \item Each client~$u$ then sends its perturbed datum~$\UsrRndData{u}$ to $v \in [n]$ with probability~$\sDist[v]$. 
    \item Each client~$u$ reports to the curator the multiset of data,~$S_u$ they receive. 
\end{enumerate}

We claim that this idealized model achieves exactly the same privacy amplification as the  shuffle model, but without relying on a central shuffler. 

\begin{figure}[!t]
	\centering
	\centering
    \begin{subfigure}{.5\textwidth}
        \centering
        \includegraphics[width=\linewidth]{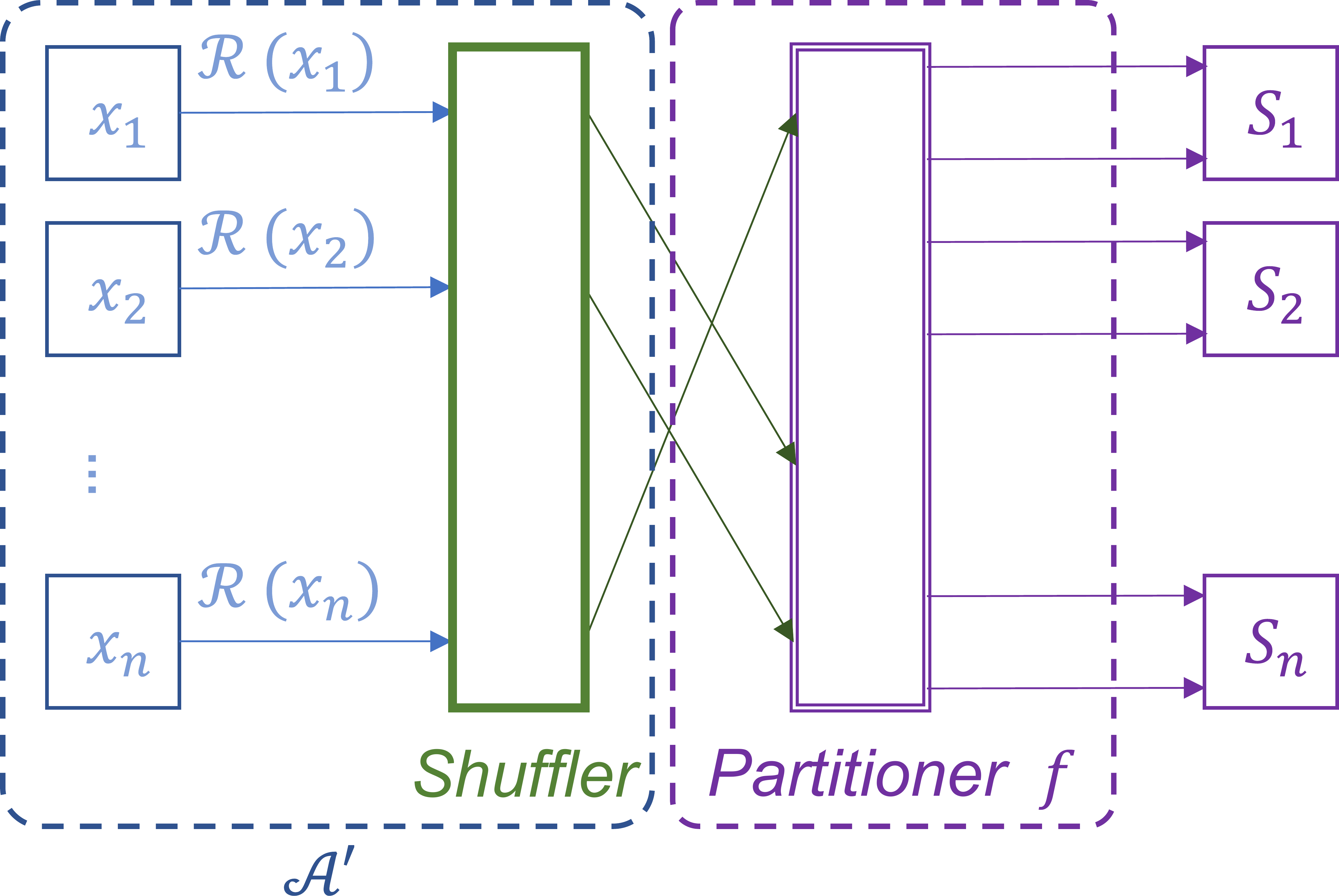}
        \caption{$\qquad$}
        \label{fig:sub1}
    \end{subfigure}%
    \begin{subfigure}{.5\textwidth}
        \centering
        \hspace{1.2cm}
        \includegraphics[width=0.75\linewidth]{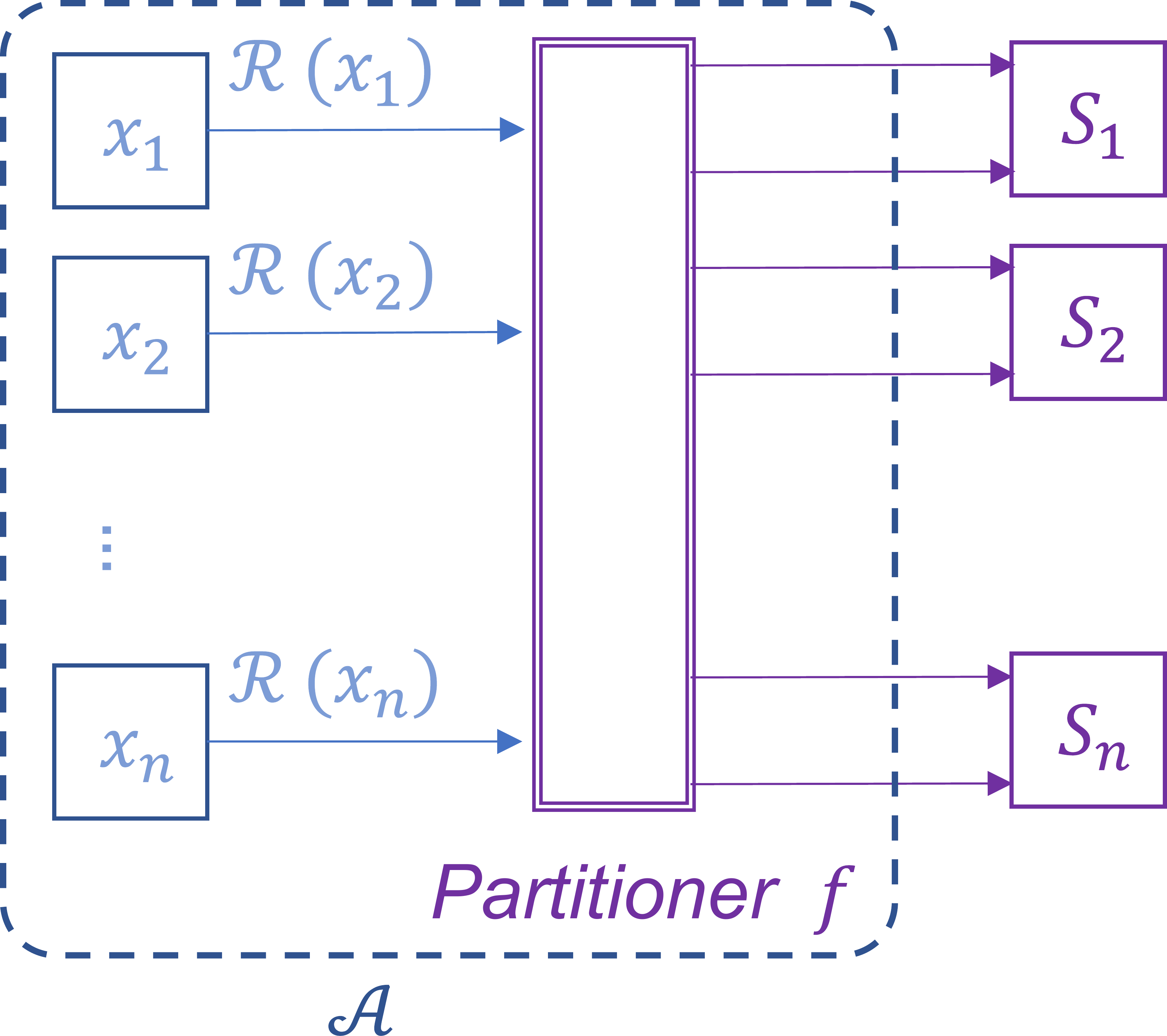}
        \caption{}
        \label{fig:sub2}
    \end{subfigure}
	\caption{
	    A pictorial illustration for the proof of Theorem~\ref{theorem: privacy amplificaiton of PerfectPartitioning}.
	    The algorithms~$f \circ \cA'$ and~$\cA$ have the same output distribution. 
	}
	\label{fig: perfect shuffling}
\end{figure}

\begin{algorithm}[!ht]
    \caption{Simulation for Infinite-Step Random Walk $\algoPerfectPartitioning{}$}
    \label{algo: inifinite time random walk}
    \begin{algorithmic}[1]
        \Require User Datum~$\UsrData{u}$.
        \State $\UsrRndData{u} \leftarrow \cR\paren{\UsrData{u}}$.
        \State Sample $v \in [n]$ according to probability distribution $\pi$.
        \State Send~$\UsrRndData{u}$ to~$v$. 
        \State Receive reports from other clients; denote the multi-set of received reports as~$\UsrSet{u}$. \\
        \Return $\UsrSet{u}$ to the curator.
    \end{algorithmic}
\end{algorithm}

\begin{theorem}[Privacy Amplification by~$\algoPerfectPartitioning$]
    \label{theorem: privacy amplificaiton of PerfectPartitioning}
    For a domain~$\cX$, let $\cR : \cX \rightarrow \cY$ be an~$\paren{\eps_0, \delta_0}$-local randomizer.
    Let~$\algoPerfectPartitioning : \cX^n \rightarrow \paren{\N^{\cY}}^n$ be the algorithm that, given a dataset $x_{1:n} \doteq \paren{x_1, \ldots, x_n} \in \cX^n$, computes~$y_1 \doteq \cR\paren{x_1}, \ldots, y_n \doteq \cR \paren{x_n}$, then randomly and independently partitions them into~$n$ multi-sets~$\UsrSet{1}, \ldots, \UsrSet{n}$, such that for each~$u \in [n]$ and each $v \in [n]$, $\P{y_u \in S_v} = \sDist[v]$. 
    Then for every~$\delta \in [0, 1]$ such that~$\eps_0 \le \log \paren{ \frac{n}{16 \log \paren{2 / \delta}} }$,~$\algoPerfectPartitioning$ is~$\paren{\eps, \delta'}$-differentially private, where~$\eps$ is as in Inequality~(\ref{ineq: privacy amplificaiton via uniform shuffling}) and~$\delta'$ is as in Equation~(\ref{equa: def of delta'}).
    
\end{theorem}

\begin{proof}
    
    Consider another algorithm~$\cA'$, which, after computing~$\UsrRndData{1} \doteq \cR \paren{x_1}, \ldots, \UsrRndData{n} \doteq \cR\paren{x_n}$, samples a uniform permutation~$\sigma$ over~$[n]$, and outputs~$y_{\sigma(1)}, \ldots, y_{\sigma(n)}$. 
    According to Fact~\ref{fact: privacy amplification via uniform shuffling},~$\cA'$ is~$\paren{\eps, \delta'}$-differentially private, where~$\eps$ is as in Inequality~(\ref{ineq: privacy amplificaiton via uniform shuffling}) and~$\delta'$ is as in Equation~(\ref{equa: def of delta'}).
    We claim that there is a post-processing algorithm~$f : \cY^n \rightarrow \paren{\N^\cY}^n$, such that~$\algoPerfectPartitioning(x_{1:n})$ is distributed identically to~$f \circ \cA'(x_{1:n}) = f \paren{y_{\sigma(1)}, \ldots, y_{\sigma(n)}}$.
    But by Fact~\ref{fact: post processing},~$f \circ \cA'(x_{1:n})$ is still~$\paren{\eps, \delta'}$-DP, which proves that~$\algoPerfectPartitioning(x_{1:n})$ is~$\paren{\eps, \delta'}$-DP. 
    
    The algorithm~$f$ is simple: for each input~$w_{1:n} \doteq \paren{w_1, \ldots, w_n}\in \cY^n$, $f$ outputs a partition~$W_1, \ldots, W_n \in \N^\cY$ of the entries of~$w_{1:n}$, such that~$\P{w_u \in W_v } = \sDist[v]$, for each~$u, v \in [n]$.
    Then~$\algoPerfectPartitioning(x_{1:n})$ is distributed exactly the same as~$f(y_1, \ldots, y_n)$. 
    On the other hand, the additional permutation step does not affect the partitioning step of~$f$, therefore~$f \paren{y_{\sigma(1)}, \ldots, y_{\sigma(n)}}$ should have the same distribution as $f(y_1, \ldots, y_n)$, which proves our claim.
    Figure~\ref{fig: perfect shuffling} includes a pictorial illustration of the proof.
\end{proof}

\subsection{Finite-Step Random Walk}
\label{subsec: finite step random walk}

In this section, we finish the proof of Theorem~\ref{theorem: privacy amplification by finite random walk}.
We compare the algorithms~$\algoPerfectPartitioning$ and~$\algoFiniteTime$.
Both algorithms compute the perturbed data~$\UsrRndData{1} \doteq \cR \paren{x_1}, \ldots, \UsrRndData{n} \doteq \cR\paren{x_n}$, and then output a partition~$S_1, \ldots, S_n$ of these data. 
The only difference is that the partitioning for~$\algoPerfectPartitioning$ is performed according to the stable distribution~$\sDist$, while the partitioning for~$\algoFiniteTime$ is performed according to the distributions of finite-step random walks.
However, according to Fact~\ref{fact: mixing time}, if the number of steps is large, then the distributions of the random walks can be sufficiently close to the stable distribution.
In such a case,~$\algoFiniteTime$ should have a similar output distribution to~$\algoPerfectPartitioning$.

\begin{lemma}
    \label{lemma: distance between finite and inifinte random walk}
    Suppose we set~$T = \paren{1 / \alpha} \ln \paren{n^{4.5} / \paren{\eps_0}}$ for Algorithm~\ref{algo: finite time random walk}.
    Then for each input dataset~$x_{1:n} \in \cX^n$, and each measurable~$Z$ in the common output space~$\paren{\N^{\cY}}^n$, it holds that 
    \begin{equation}
        \label{ineq: distance between finite and inifinte random walk}
        e^{-\eps_0 / \paren{2n}} \le 
        \frac{
            \P{ \algoFiniteTime (x_{1:n}) \in Z }
        }{
            \P{ \algoPerfectPartitioning (x_{1:n}) \in Z }
        } 
        \le e^{\eps_0 / \paren{2n}}.
    \end{equation}
\end{lemma}
To save space, the proof of Lemma~\ref{lemma: distance between finite and inifinte random walk} is  in Appendix~\ref{appendix: proof for sec: proof of privacy amplification};
we complete the proof for Theorem~\ref{theorem: privacy amplification by finite random walk}. 
\begin{proof}[Proof of Theorem~\ref{theorem: privacy amplification by finite random walk}.]
    Based on Theorem~\ref{theorem: privacy amplificaiton of PerfectPartitioning} and Lemma~\ref{lemma: distance between finite and inifinte random walk}, for each pair of neighboring datasets~$x_{1:n}, x_{1:n}' \in \cX^n$ and each measurable~$Z$ in common output space~$\paren{\N^{\cY}}^n$, we have
    \begin{align*}
         \P{ \algoFiniteTime (x_{1:n}) \in Z }
            &\le e^{\eps_0 / \paren{2n}} \P{ \algoPerfectPartitioning (x_{1:n}) \in Z } \\
            &\le e^{\eps_0 / \paren{2n}} \left( e^\eps \P{ \algoPerfectPartitioning (x_{1:n}') \in Z } + \delta' \right)  \\
            &\le e^{\eps_0 / \paren{2n}} \left( e^\eps e^{\eps_0 / \paren{2n}} \P{ \algoFiniteTime (x_{1:n}') \in Z } + \delta' \right)  \\
            &= e^{ \eps + \eps_0 / n} \P{ \algoFiniteTime (x_{1:n}') \in Z } + e^{\eps_0 / \paren{2n}} \delta'\,, 
    \end{align*}
    where~$\eps$ is as in Inequality~(\ref{ineq: privacy amplificaiton via uniform shuffling}) and~$\delta'$ is as in Equation~(\ref{equa: def of delta'}).
    This finishes the proof. 
\end{proof}

\section{Subsampling and Shuffling}
\label{sec: subsampling}

\vspace{-2mm}
In this section, we show how subsampling further enhances the privacy guarantee of shuffling. 
We observe that our bound still applies, if we replace the network shuffle model with the shuffle model, i.e., if we combine Poisson subsampling with the shuffle model.

Since there is no trusted sampler, the subsampling is performed in a fully decentralized manner via \emph{Poisson subsampling}, where each client is sampled independently with some specified probability~$p$.
We refer to the network shuffling protocol that incorporates subsampling as~$\algoSampleThenWalk$.
Algorithm~\ref{algo: sample then random walk} describes the client-side algorithm of~$\algoSampleThenWalk$, which 
 is almost the same as Algorithm~\ref{algo: finite time random walk}, except that each client now only reports its randomized datum with probability~$p$~(line~\ref{line: algo: sample then random walk: sample step}, Algorithm~\ref{algo: sample then random walk}).

\begin{algorithm}[!ht]
    \caption{Client-Side Protocol for~$\algoSampleThenWalk$}
    \label{algo: sample then random walk}
    \begin{algorithmic}[1]
        \Require User Datum~$\UsrData{u} \in \cD$; Probability~$p \in \paren{0, 1}$.
        \State Same as Algorithm~\ref{algo: finite time random walk}, except that Line~\ref{line: algo: finite time random walk: set report set} is replaced with
        \Statex $\qquad \UsrSet{u} \leftarrow 
            \begin{cases}
                \{ \UsrRndData{u} \},   &\text{w.p.}\, p, \\
                \emptyset,              &\text{w.p.}\, 1-p. 
            \end{cases}$
        \label{line: algo: sample then random walk: sample step}
    \end{algorithmic}
\end{algorithm}

We now analyze the privacy amplification bound for~$\algoSampleThenWalk$ formally. 
For simplicity, we assume that $p = k / n$ for some~$k \in [n]$, although our result also holds if we consider real values of~$p$.

\begin{theorem}[Privacy Amplification by~$\algoSampleThenWalk$]
    \label{theorem: privacy amplificaiton by sampling and random walk}
    Let $\cR : \cX \rightarrow \cY$ be an arbitrary $\paren{\eps_0, \delta_0}$-differentially private local randomizer.
    Let~$\algoSampleThenWalk$ be the algorithm 
    as described by Algorithm~\ref{algo: sample then random walk} for~$T = \paren{1 / \alpha} \ln \paren{n^{4.5} / \paren{\eps_0}}$ rounds of random walks, where~$\alpha$ is the spectral gap of~$G$ (Definition~\ref{def: spectral gap}). 
    Then for every~$\delta \in [0, 1]$ such that~$\eps_0 \le \log \paren{ \frac{k - n \lambda\paren{p}}{16 \log \paren{2 / \delta}} }$,~$\algoSampleThenWalk$ is~$\paren{\eps', \delta'}$-differentially private, where~$\delta' = \delta +  \paren{k / n + \lambda\paren{p}} e^{\eps_0 / \paren{2n}} \paren{\delta + \paren{e^\eps + 1}\paren{1 + e^{-\eps_0} / 2} \paren{k + n \lambda\paren{p}} \delta_0}$ 
    and 
    \vspace{-1mm}
    \begin{equation}
        \label{ineq: privacy amplificaiton via sample then random walk}
        \eps' \le 
        \frac{\eps_0}{n} + 
        \ln \PAREN{1 + \frac{e^{\eps_0} - 1}{e^{\eps_0} + 1} \PAREN{
                \sqrt{\frac{k}{n} + \lambda \paren{ p }} \frac{
                    8\sqrt{e^{\eps_0} \log \paren{4 / \delta}}
                }{
                    \sqrt{n}
                } + 
                \frac{
                    8 e^{\eps_0}
                }{
                    n
                }
            }
        }\,, 
    \end{equation}
    where~$\lambda\paren{p} \doteq \sqrt{ \paren{2 p (1 - p) / n} \ln \paren{2 / \delta}} + \paren{2 / \paren{3n}} \ln \paren{2 / \delta} \in o(1)$.
    In particular, when~$\cR$ is~$\eps_0$-differentially private, then $\algoSampleThenWalk$ is~$(\eps', \delta + \paren{k / n + \lambda \paren{p}} e^{\eps_0 / \paren{2n}} \delta )$-differentially private. 
\end{theorem}

{\bf Building Blocks.} 
Our proof relies on the following result of privacy amplification of subsampling without
replacement proved by~\citeauthor{BalleBG18}~\citep{BalleBG18}.

\begin{fact}[Privacy Amplification by Subsampling~(Theorem 9~\citep{BalleBG18})]
    \label{fact: privacy amplification by subsampling}
    Let~$\cA: \cX^\ell \rightarrow Z$ be an~$\paren{\eps, \delta}$-differentially private algorithm, and~$\cA': \cX^n \rightarrow Z$ be the algorithm that given a dataset~$x_{1:n} \doteq \paren{x_1, \ldots, x_n} \in \cX^n$, samples a subset of~$\ell$ elements from~$x_{1:n}$ uniformly at random and without replacement, then applies~$\cA$ on the sampled subsets.
    It holds that~$\cA'$ is~$\paren{\eps', \paren{\ell / n} \delta}$-differentially private, where
    \vspace{-2mm}
    \begin{equation}
        \eps' = \ln \paren{1 + \paren{\ell / n} \paren{e^\eps - 1}}\,.
    \end{equation}
\end{fact}

There is another technical lemma we require for establishing the proof of Theorem~\ref{theorem: privacy amplificaiton by sampling and random walk}. 
Let~$\cC \subseteq [n]$ be a subset of~$\ell$ clients.
Suppose we modify the client protocol for~$\algoFiniteTime$ (Algorithm~\ref{algo: finite time random walk}) by replacing Line~\ref{line: algo: finite time random walk: set report set} with 
$\UsrSet{u} \leftarrow 
\begin{cases}
            \{ \UsrRndData{u} \},   &\forall u \in \cC, \\
            \emptyset,              &\text{otherwise}. 
\end{cases}$, 
such that only clients in~$\cC$ report their perturbed elements.
Let the resulting algorithm be~$\algoFiniteTime^{\cC}$. 
Its privacy guarantee is stated as follows. 

\begin{lemma} 
    \label{lemma: privacy amplification by partial shuffle}
    For every~$\delta \in [0, 1]$ such that~$\eps_0 \le \log \left( \frac{\ell}{16 \log \paren{2 / \delta_0}}\right)$,
    ~$\algoFiniteTime^{\cC}$ is~$\paren{\eps, \delta''}$-DP, where 
    $
        \eps \le 
        \frac{\eps_0 }{n} + \log 
        \PAREN{1 + \frac{e^{\eps_0} - 1}{e^{\eps_0} + 1}                  \PAREN{
                \frac{
                    8 \sqrt{e^{\eps_0} \log \paren{4 / \delta}}
                }{
                    \sqrt{\ell}
                } + 
                \frac{
                    8 e^{\eps_0}
                }{
                    \ell
                }
            }
        }
    $ and~$\delta'' \doteq e^{\eps_0 / \paren{2n}} \paren{\delta + \paren{e^\eps + 1}\paren{1 + e^{-\eps_0} / 2} \ell \delta_0}$.
\end{lemma}

The proof for the lemma is almost identical to that for Theorem~\ref{theorem: privacy amplification by finite random walk},
with a slight modification. 
For completeness, the proof is in Appendix~\ref{appendix: proof for sec: subsampling}.
Next, we sketch a proof of Theorem~\ref{theorem: privacy amplificaiton by sampling and random walk}. 

\begin{proof}[\bf Sketch Proof of Theorem~\ref{theorem: privacy amplificaiton by sampling and random walk}]
    Let $U$ be the set of clients sampled by~$\algoSampleThenWalk$.
    For each~$\ell \in \IntSet{0}{n}$, let~$\binom{[n]}{\ell}$ be the collection of subsets of~$[n]$ with size~$\ell$. 
    Since each client is sampled independently with equal probability, conditioned on~$|U| = \ell$, 
    $U$ can be viewed as a random element sampled uniformly from~$\binom{[n]}{\ell}$. 
    Therefore, we can view~$\algoSampleThenWalk$ as~$\algoFiniteTime^{U}$. 
    By Fact~\ref{fact: privacy amplification by subsampling} and Lemma~\ref{lemma: privacy amplification by partial shuffle}, we can show that~$\algoFiniteTime^{U}$ is~$\paren{\eps_l, \delta_\ell}$-DP, where 
    $
        \eps_\ell
            \le \frac{\eps_0}{n} + \ln \PAREN{
                    1 + \PAREN{
                            \frac{e^{\eps_0} - 1}{e^{\eps_0} + 1} \PAREN{
                            \sqrt{\frac{\ell}{n}}
                            \frac{
                                8 \sqrt{e^{\eps_0} \log \paren{4 / \delta}}
                            }{
                                \sqrt{n}
                            }
                            + \frac{
                                8 e^{\eps_0}
                            }{
                                n
                            }
                        }
                } 
            }
    $
    and
    $
        \delta_\ell = \paren{\ell / n } e^{\eps_0 / \paren{2n}} \paren{\delta + \paren{e^\eps + 1}\paren{1 + e^{-\eps_0} / 2} \ell \delta_0}.
    $
    Note that both~$\eps_\ell$ and~$\delta_\ell$ are increasing functions of~$\ell$. 
    Next, via a Chernoff-style concentration inequality, we can prove that, the size of~$U$ will concentrate around~$np$. 
    Therefore, the privacy guarantee is roughly~$\paren{\eps_\ell, \delta_\ell}$.
\end{proof}

\vspace{-4mm}
{\bf Comparison with Previous Approach.}
Subsampling was also studied by~\citeauthor{Liew2022}~\citep{Liew2022} for the network shuffle model. 
However, their subsampling strategy~\citep{Liew2022} is different: after the random walks for the perturbed data stop, each client~$u$ only samples and reports one datum from the multi-set~$S_u$ of data they receives. 
This strategy has no control over the number of reports returned to the curator. 
Indeed, it can be proven that, there exists an undirected graph, in which there can be some client~$u$ who receives~$\Omega(n)$ perturbed data in expectation (after sufficient number of random walks).  
In such a case, client~$u$ will drop~$\Omega(n)$ reports. 
Further, their privacy guarantee is~$\paren{O\paren{ e^{\eps_0} \paren{e^{\eps_0} - 1} \sqrt{ \paren{(1 - \alpha)^{2T} + \sum_{i \in [n]} \paren{\sDist[i]}^2} \ln \paren{1 / \delta} } }, \delta}$, which is not as good as ours.

\vspace{-2mm}
\section{Related Works} \label{sec: review}

\vspace{-2mm}
{\bf Shuffle Model.} 
Shuffling is a well known technique for enhancing the privacy guarantees of the locally randomized reports generated by the clients. 
The early work includes the \textsc{Prochlo} system proposed and implemented by~\citeauthor{BittauEMMRLRKTS17}~\citep{BittauEMMRLRKTS17} 
which applies shuffling to provide anonymity of local reports. 
Since~\citep{BittauEMMRLRKTS17} did not have a formal privacy analysis,
a series of works endeavoured to obtain tight privacy guarantee for the model.  
Later,~\citeauthor{EFMRTT19}~\citep{EFMRTT19} studied the mechanism where the clients' data are shuffled before applying the local randomizers.
This work allows the local randomizers to be chosen adaptively: the output of a local randomizer depends not only on the current input, but also possibly on previous outputs of the local randomizers. 
\citeauthor{EFMRTT19}~\citep{EFMRTT19} prove that if the local randomizers are~$\eps_0$-DP, then the output of the mechanism is $\paren{ O\paren{e^{2\eps_0} \paren{e^{\eps_0} - 1} \sqrt{\paren{1 / n} \ln \paren{1 / \delta}}}, \delta}$-DP. 
Observe that, if the same local randomizer (with independent randomness) is applied to all data, then applying the local randomizer to the shuffled data is distributed identically to shuffling the data perturbed by the local randomizer.
In this case, the privacy analysis still holds. 
The bound was tightened to~$\paren{ O\paren{e^{3\eps_0 / 2} \paren{e^{\eps_0} - 1} \sqrt{\paren{1 / n} \ln \paren{1 / \delta}}}, \delta}$ by~\citeauthor{BalleKMTT20}~\citep{BalleKMTT20}.
They also studied the privacy amplification for approximate differentially private local randomizer: if the local randomizers are~$\paren{\eps_0, \delta_0}$-DP, then they showed that the output of the mechanism is~$\paren{\eps', \delta'}$-DP, where~$\eps' \in O\paren{e^{12\eps_0} \paren{e^{8\eps_0} - 1} \sqrt{\paren{1 / n} \ln \paren{1 / \delta}}}$ and~$\delta' \in O\paren{ \delta + n \paren{e^{\eps'} + 1} \delta_0}$. 
On the other hand, if the clients share the same local algorithm (with independent randomness) and if the shuffling is applied to locally randomized data, \citeauthor{BalleBGN19}~\citep{BalleBGN19} show that the mechanism is~$\paren{ O \paren{ \min \set{ \eps_0, 1} \cdot e^{\eps_0} \sqrt{\paren{1 / n} \ln \paren{1 / \delta}}}, \delta}$-DP. 
When the local algorithm outputs a binary randomized response, \citeauthor{CheuSUZZ19}~\citep{CheuSUZZ19} proves the privacy guarantee can be further improved to~$\paren{ O \paren{  \sqrt{\paren{e^{\eps_0} / n} \ln \paren{1 / \delta}}}, \delta}$. 
This line of research culminated with the work by~\citeauthor{FeldmanMT21}~\citep{FeldmanMT21} who followed the mechanism studied by~\citeauthor{EFMRTT19}~\citep{EFMRTT19}, and proved if the local randomizers are~$\paren{\eps_0, \delta_0}$-DP, then the output of the model is~$\paren{\eps', \delta'}$-DP, where~$\eps' \in O\paren{ \paren{1 - e^{-\eps_0}}\sqrt{\paren{e^{\eps_0} / n} \ln \paren{1 / \delta}}}$ and~$\delta' \in O\paren{ \delta + n \paren{e^{\eps'} + 1} \paren{1 + e^{-\eps_0} / 2}\delta_0}$. This is the state-of-the-art privacy amplification bound for shuffle model.

{\bf Subsampling.} 
Another key technique for enhancing privacy guarantee is subsampling, which runs a differentially private algorithm on a randomly sampled subset of clients instead of on the whole population. 
\citeauthor{BalleBG18}~\citep{BalleBG18} studied several subsampling techniques, namely Poisson sampling, sampling with/without replacement, based on different definitions of neighboring datasets. 
It is worth noting that, a trusted central curator is assumed in \citep{BalleBG18}, and that privacy amplification bound for Poisson subsampling w.r.t.~neighboring datasets defined in terms of substitution relationship (the definition of neighboring datasets applied in our paper) is cumbersome.
Since it is not known how this bound can be applied directly to the model studied in our paper,
our analysis is based on a reduction to subsampling without replacement~\citep{BalleBG18}. 
\citeauthor{BalleKMTT20}~\citep{BalleKMTT20} also studied subsampling for differentially private stochastic gradient descent (DP-SGD) optimization in the setting of Federated Learning. 
In their work~\citep{BalleKMTT20}, the client decides to participate in the training locally and independently, and a privacy amplification bound similar to, but not as tight as, our bound in Theorem~\ref{theorem: privacy amplificaiton by sampling and random walk}, is presented. 
Further, the work by~\citeauthor{BalleKMTT20}~\citep{BalleKMTT20} 1) requires a trusted central server, which will learn the client identities, as compared to the fully decentralized algorithm studied in our work; 2) is dedicated for (DP-SGD) optimization which allows dummy updates.

\newpage
\bibliographystyle{IEEEtranN}
\bibliography{reference}

\newpage
\appendix
\section{Concentration Inequality}

\begin{fact}[Bernstein’s Inequality~\citep{AMS09}] \label{fact: bernstein} 
    Let $X_1, \ldots, X_n$ be independent real-valued random variables such that $|X_i| \le c$ with probability one. 
    Let $S_n = \sum_{i \in [n]} X_i$ and $\Var{S_n} = \sum_{ i \in [n] } \Var{X_i^2}$. 
    Then for all $\beta \in (0, 1)$, with probability at least $1 - \beta$, 
    $$
        \left| S_n - \E{S_n} \right| \le \sqrt{ 2 \Var{S_n} \ln \frac{2}{\beta} } + \frac{2 c \ln \paren{2 / \beta}}{ 3}\,,
    $$ 
\end{fact}

\section{Proofs for Section~\ref{sec: proof of privacy amplification}}
\label{appendix: proof for sec: proof of privacy amplification}

\begin{proof}[Proof of Lemma~\ref{lemma: distance between finite and inifinte random walk}]

    Define the function~$g: \cY^n \times [n]^n \rightarrow \paren{\N^{\cY}}^n$, which when given input~$w_{1:n} = \paren{w_1, \ldots, w_n} \in \cY^n$ and~$\ell_{1:n} = \paren{\ell_1, \ldots, \ell_n} \in [n]^n$, outputs a partition~$S_1, \ldots, S_n$ of the entries of~$w_{1:n}$, such that~$S_i \doteq \{ w_u : \ell_u = i \}$ for each~$i \in [n]$.
    
    Let~$\vecp{u}$ be the probability distribution of a~$T$-step random walk staring at vertex~$u$. 
    According to Fact~\ref{fact: mixing time}, when~$T = \paren{1 / \alpha} \ln \paren{n^{4.5} / \paren{\eps_0}}$, 
    \begin{equation}
        \label{ineq: deviation of pu}
        \norm{\vecp{u} - \sDist}_1 \le \frac{\eps_0}{n^4}, \quad \forall u \in [n]. 
    \end{equation}
    
    Given input~$x_1, \ldots, x_n$, define~$\vec{Y} \doteq \paren{\cR^{(1)}\paren{x_1}, , \ldots, \cR^{(n)}\paren{x_n}}$. 
    For each~$u \in [n]$, define the random variable~$I_u$ over~$[n]$, such that~$\P{I_u = i} = \vecp{u}[i]$ for each~$i \in [n]$. 
    Let~$\vec{I}$ be shorthand for~$\paren{I_1, \ldots, I_n}$.
    It is easy to see that~$g \paren{\vec{Y}, \vec{I}}$ has exactly the same distribution as~$\algoFiniteTime$. 
    
    Similarly, for each~$u \in [n]$, define the random variable~$J_u$ over~$[n]$, such that~$\P{J_u = i} = \sDist[i]$ for each~$i \in [n]$. 
    Let~$\vec{J}$ be shorthand for~$\paren{J_1, \ldots, J_n}$.
    Then~$g \PAREN{\vec{Y}, \vec{J}}$ has exactly the same distribution as~$\algoPerfectPartitioning$. 
    
    In order to prove Inequality~(\ref{ineq: distance between finite and inifinte random walk}), it suffices to prove that for each measurable subset~$Z \subseteq \paren{\N^{\cY}}^n$, it holds that 
    \begin{equation*}
        e^{-\frac{\eps_0}{2n}}
        \le 
        \frac{\P{ g \paren{\vec{Y}, \vec{J}} \in Z }}{\P{ g \paren{\vec{Y}, \vec{I}} \in Z }} 
        \le 
        e^{\frac{\eps_0}{2n}},
        \qquad
        \text{or}
        \qquad 
        e^{-\frac{\eps_0}{2n}} 
        \le 
        \frac{\P{ \paren{\vec{Y}, \vec{J}} \in g^{-1} \paren{Z} } }{\P{ \PAREN{\vec{Y}, \vec{I}} \in g^{-1} \paren{Z} }}
        \le 
        e^{\frac{\eps_0}{2n}},
    \end{equation*}
    where~$g^{-1} \paren{Z} \doteq \set{ \PAREN{w_{1:n}, \ell_{1:n} } \in \cY^n \times [n]^n : g \PAREN{w_{1:n}, \ell_{1:n}} \in Z}$ is the preimage of~$Z$ under~$g$.
    To prove the inequalities on the RHS, we claim that for each~$w_{1:n} \in \cY^n$, and each~$\ell_{1:n} \in [n]^n$,
    \begin{equation*}
        e^{-\frac{\eps_0}{2n}}
        \le 
        \frac{\P{ \paren{\vec{Y}, \vec{J}} = \PAREN{w_{1:n}, \ell_{1:n} } }}{\P{ \paren{\vec{Y}, \vec{I}} = \PAREN{w_{1:n}, \ell_{1:n}} }} 
        \le 
        e^{\frac{\eps_0}{2n}}, 
        \quad
        \text{or}
        \quad 
        e^{-\frac{\eps_0}{2n}}
        \le 
        \frac{\P{ \vec{J} = \ell_{1:n} \mid \vec{Y} = w_{1:n} }}{\P{ \vec{I} = \ell_{1:n} \mid \vec{Y} = w_{1:n} }} 
        \le 
        e^{\frac{\eps_0}{2n}}.
    \end{equation*}
    The inequalities on the LHS and RHS are equivalent, since both~$\paren{\vec{Y}, \vec{I}}$ and~$\PAREN{\vec{Y}, \vec{J}}$ have the same distribution for~$\vec{Y}$, and~$\P{ \paren{\vec{Y}, \vec{J}} = \PAREN{w_{1:n}, \ell_{1:n} } } = \P{ \vec{J} = \ell_{1:n} \mid \vec{Y} = w_{1:n} } \P{\vec{Y} = w_{1:n}}$ and~$\P{ \paren{\vec{Y}, \vec{I}} = \PAREN{w_{1:n}, \ell_{1:n} } } = \P{ \vec{I} = \ell_{1:n} \mid \vec{Y} = w_{1:n} } \P{\vec{Y} = w_{1:n}}$. 
    It remains to verify the inequalities on the RHS.
    By definitions, we have 
    \begin{align*}
        \P{ \vec{I} = \ell_{1:n} \mid \vec{Y} = w_{1:n} } = \prod_{u \in [n]} \sDist\bracket{\ell_u}, \qquad
        \P{ \vec{J} = \ell_{1:n} \mid \vec{Y} = w_{1:n} } = \prod_{u \in [n]} \vecp{u}\bracket{\ell_u}. 
    \end{align*}
    Therefore, 
    \begin{align*}
        \frac{
            \P{ \vec{J} = \ell_{1:n} \mid \vec{Y} = w_{1:n} }
        }{
            \P{ \vec{I} = \ell_{1:n} \mid \vec{Y} = w_{1:n} }
        }
        &= \frac{
            \prod_{u \in [n]} \vecp{u}\bracket{\ell_u}
        }{
            \prod_{u \in [n]} \sDist\bracket{\ell_u}
        }
        \le \prod_{u \in [n]} \frac{
            \sDist\bracket{\ell_u} + \eps_0  / n^4
        }{
            \sDist\bracket{\ell_u}
        } 
        \\
        &\le \prod_{u \in [n]} \exp \PAREN{
            \frac{
                \eps_0 / n^4
            }{
                \sDist\bracket{\ell_u}
            }
        } 
        \le \prod_{u \in [n]} \exp \PAREN{\frac{\eps_0}{n^2}} 
        = \exp \PAREN{\frac{\eps_0}{2n}},
    \end{align*}
    where the first inequality holds since~$\big| \vecp{u}\big[\ell_u\big] - \sDist\big[\ell_u\big]\big| \le \norm{\vecp{u} - \sDist}_1 \le \eps_0 / n^4$, 
    the second inequality follows from that~$1 + x \le e^x$ for all~$x \in \R$, 
    the third inequality follows from that~$\sDist\big[\ell_u\big] = d_v / \paren{2 |E|} \ge 2 / \paren{n \paren{n - 2}}$.
    Similarly, 
    \begin{align*}
        \frac{
            \P{ \vec{J} = \ell_{1:n} \mid \vec{Y} = w_{1:n} }
        }{
            \P{ \vec{I} = \ell_{1:n} \mid \vec{Y} = w_{1:n} }
        }
        &= \frac{
            \prod_{u \in [n]} \vecp{u}\bracket{\ell_u}
        }{
            \prod_{u \in [n]} \sDist\bracket{\ell_u}
        }
        \ge \prod_{u \in [n]} \frac{
            \sDist\bracket{\ell_u} - \eps_0 / n^4
        }{
            \sDist\bracket{\ell_u}
        } 
        \\
        &\ge \prod_{u \in [n]} \exp \PAREN{-2 \cdot 
            \frac{
                \eps_0 / n^4
            }{
                \sDist\bracket{\ell_u}
            }
        } 
        \ge \prod_{u \in [n]} \exp \PAREN{ -2 \cdot \frac{\eps_0}{n^2}} 
        = \exp \PAREN{-2 \cdot \frac{\eps_0}{2n}},
    \end{align*}
    where the second inequality follows from that~$1 - x \ge e^{-2x}$ for each~$x \in [0, 3 / 4]$.
\end{proof}

\section{Proofs for Section~\ref{sec: subsampling}}
\label{appendix: proof for sec: subsampling}

\begin{proof}[\bf Proof of Theorem~\ref{theorem: privacy amplificaiton by sampling and random walk}]
    Let $U$ be the set of clients sampled by~$\algoSampleThenWalk$.
    For each~$\ell \in \IntSet{0}{n}$, let~$\binom{[n]}{\ell}$ be the collection of subsets of~$[n]$ with size~$\ell$. 
    Since each client is sampled independently with equal probability, conditioned on~$|U| = \ell$, 
    $U$ can be viewed as a random element sampled uniformly from~$\binom{[n]}{\ell}$. 
    Therefore, we can view~$\algoSampleThenWalk$ as~$\algoFiniteTime^{U}$. 
    By Fact~\ref{fact: privacy amplification by subsampling} and Lemma~\ref{lemma: privacy amplification by partial shuffle},
    for every~$\delta \in [0, 1]$ such that~$\eps_0 \le \log \left( \frac{\ell}{16 \log \paren{2 / \delta_0}}\right)$,~$\algoFiniteTime^{U}$ is~$\paren{\ell_l, \delta_\ell}$-DP, where 
    \begin{align*}
        \label{equa: def of eps ell}
        \eps_\ell 
            &= \ln \PAREN{ 
                1 + \frac{\ell}{n} \PAREN{ 
                    e^{\frac{\eps_0 }{n}} \PAREN{
                    1 + \frac{e^{\eps_0} - 1}{e^{\eps_0} + 1}                  \PAREN{
                            \frac{
                                8 \sqrt{e^{\eps_0} \log \paren{4 / \delta}}
                            }{
                                \sqrt{\ell}
                            } + 
                            \frac{
                                8 e^{\eps_0}
                            }{
                                \ell
                            }
                        }
                    } 
                    - 1 
                }
            }
            \\
            &= \ln \PAREN{
                    1 + \PAREN{
                            \frac{e^{\eps_0} - 1}{e^{\eps_0} + 1} \PAREN{
                            \sqrt{\frac{\ell}{n}}
                            \frac{
                                8 \sqrt{e^{\eps_0} \log \paren{4 / \delta}}
                            }{
                                \sqrt{n}
                            }
                            + \frac{
                                8 e^{\eps_0}
                            }{
                                n
                            }
                        }
                } e^{\frac{\eps_0}{n}}
                + \frac{\ell}{n} \cdot  \PAREN{e^{\frac{\eps_0}{n}} - 1}
            } \\
            &= \ln \PAREN{
                    \PAREN{
                            \frac{e^{\eps_0} - 1}{e^{\eps_0} + 1} \PAREN{
                            \sqrt{\frac{\ell}{n}}
                            \frac{
                                8 \sqrt{e^{\eps_0} \log \paren{4 / \delta}}
                            }{
                                \sqrt{n}
                            }
                            + \frac{
                                8 e^{\eps_0}
                            }{
                                n
                            }
                        }
                } e^{\frac{\eps_0}{n}}
                + \frac{\ell}{n} \cdot e^{\frac{\eps_0}{n}} 
                + \frac{n - \ell}{n}
            } \\
            &\le \ln \PAREN{
                    \PAREN{
                            \frac{e^{\eps_0} - 1}{e^{\eps_0} + 1} \PAREN{
                            \sqrt{\frac{\ell}{n}}
                            \frac{
                                8 \sqrt{e^{\eps_0} \log \paren{4 / \delta}}
                            }{
                                \sqrt{n}
                            }
                            + \frac{
                                8 e^{\eps_0}
                            }{
                                n
                            }
                        }
                } e^{\frac{\eps_0}{n}}
                + e^{\frac{\eps_0}{n}}
            } \\
            &\le \frac{\eps_0}{n} + \ln \PAREN{
                    1 + \PAREN{
                            \frac{e^{\eps_0} - 1}{e^{\eps_0} + 1} \PAREN{
                            \sqrt{\frac{\ell}{n}}
                            \frac{
                                8 \sqrt{e^{\eps_0} \log \paren{4 / \delta}}
                            }{
                                \sqrt{n}
                            }
                            + \frac{
                                8 e^{\eps_0}
                            }{
                                n
                            }
                        }
                } 
            }
    \end{align*}
    and
    \begin{equation}
        \label{equa: def of delta ell}
        \delta_\ell = \paren{\ell / n } e^{\eps_0 / \paren{2n}} \paren{\delta + \paren{e^\eps + 1}\paren{1 + e^{-\eps_0} / 2} \ell \delta_0}.
    \end{equation}

    Next, we pick~$\Ub \doteq \min \set{n,  n p + \sqrt{ 2 n p (1 - p) \ln \paren{2 / \delta}} + \paren{2 / 3} \ln \paren{2 / \delta} }$ and~$\Lb \doteq \max \set{0,  n p - \sqrt{ 2 n p (1 - p) \ln \paren{2 / \delta}} - \paren{2 / 3} \ln \paren{2 / \delta} }$. 
    Via Bernstein’s Inequality (Fact~\ref{fact: bernstein}), we have~$\P{|U| \notin \IntSet{\Lb}{\Ub} } \le \delta$. 
    Therefore, for each measurable subset~$Z \in \paren{\N^{\cY}}$ in the output space of~$\algoSampleThenWalk$, and each pair of neighboring datasets~$x_{1:n}$ and~$x_{1:n}'$, 
    and for every~$\delta \in [0, 1]$ such that~$\eps_0 \le \log \left( \frac{\Lb}{16 \log \paren{2 / \delta_0}}\right)$
    it holds that 
    \begin{align*}
        \P{ \algoSampleThenWalk \paren{x_{1:n}} \in Z } 
            &= \sum_{\ell \in \IntSet{0}{n} } \P{ \algoFiniteTime^{U} \paren{x_{1:n}} \in Z \middle\vert  |U| = \ell  } \cdot \P{|U| = \ell} \\
            &\le \sum_{\ell \in \IntSet{\Lb}{\Ub} } \P{ \algoFiniteTime^{U} \paren{x_{1:n}} \in Z \middle\vert  |U| = \ell  } \P{|U| = \ell } + \delta \\
            &\le \sum_{\ell \in \IntSet{\Lb}{\Ub} } \PAREN{ e^{\eps_\ell} \P{ \algoFiniteTime^{U} \paren{x_{1:n}'} \in Z \middle\vert  |U| = \ell  } + \delta_\ell } \P{|U| = \ell } + \delta \\
            &\le e^{\eps_\Ub} \P{ \algoSampleThenWalk \paren{x_{1:n}} \in Z } + \delta_\Ub + \delta.
    \end{align*}
    By the definition of~$\lambda\paren{p}$ and replacing~$p$ with $k / n$, we get
    $$
        \delta_\Ub \le \paren{k / n + \lambda\paren{p}} e^{\eps_0 / \paren{2n}} \paren{\delta + \paren{e^\eps + 1}\paren{1 + e^{-\eps_0} / 2} \paren{k + n \lambda\paren{p}} \delta_0},
    $$
    and
    $$
        \eps_\Ub \le \ln \PAREN{1 + \frac{e^{\eps_0} - 1}{e^{\eps_0} + 1} \PAREN{
                \sqrt{\frac{k}{n} + \lambda\paren{p}} \frac{
                    8 \sqrt{e^{\eps_0} \log \paren{4 / \delta}}
                }{
                    \sqrt{n}
                } + 
                \frac{
                    8 e^{\eps_0}
                }{
                    n
                }
            }
        + \frac{ e^{\eps_0} }{n}}, 
    $$
    which finishes the proof.
\end{proof}

\begin{proof}[Proof of Lemma~\ref{lemma: privacy amplification by partial shuffle}]
    First consider the modification of Algorithm~\ref{algo: inifinite time random walk}, such that only clients in~$\cC \subseteq [n]$ send their perturbed elements to randomly chosen~$v$ according to probability distribution~$\sDist$. 
    Denote the resulting algorithm as~$\algoPerfectPartitioning^{\cC}$. 
    
    Based on similar proof as Theorem~\ref{theorem: privacy amplificaiton of PerfectPartitioning}, the output of~$\algoPerfectPartitioning^{\cC}$ can be viewed as a post-processing of the shuffling of the perturbed elements~$\set{ \cR(x_u) : u \in \cC}$, where~$\cR$ is~$\paren{\eps_0, \delta_0}$-differentially private. 
    Since that~$\card{\cC} = \ell$, according to Fact~\ref{fact: privacy amplification via uniform shuffling}, for every~$\delta \in [0, 1]$ such that~$\eps_0 \le \log \left( \frac{\ell}{16 \log \paren{2 / \delta_0}}\right)$,~$\algoPerfectPartitioning^{\cC}$ is~$\paren{\eps, \delta'}$-differentially private, where
    \begin{equation}
        \eps \le \log 
        \PAREN{1 + \frac{e^{\eps_0} - 1}{e^{\eps_0} + 1}                  \PAREN{
                \frac{
                    8 \sqrt{e^{\eps_0} \log \paren{4 / \delta}}
                }{
                    \sqrt{\ell}
                } + 
                \frac{
                    8 e^{\eps_0}
                }{
                    \ell
                }
            } 
        }\,.
    \end{equation}
    and
    \begin{equation}
        \delta' = \delta + \paren{e^\eps + 1}\paren{1 + e^{-\eps_0} / 2} \ell \delta_0. 
    \end{equation}
    
    Next, consider the algorithm~$\algoFiniteTime^{\cC}$, which is obtained by modifying the client protocol for~$\algoFiniteTime$ (Algorithm~\ref{algo: finite time random walk}) by replacing Line~\ref{line: algo: finite time random walk: set report set} with 
    $\UsrSet{u} \leftarrow 
    \begin{cases}
                \{ \UsrRndData{u} \},   &\forall u \in \cC, \\
                \emptyset,              &\text{otherwise}. 
    \end{cases}$, 
    such that only clients in~$\cC$ report their perturbed elements.
    Following similar proof as Lemma~\ref{lemma: distance between finite and inifinte random walk}, suppose we set~$T = \paren{1 / \alpha} \ln \paren{n^{4.5} / \paren{\eps_0}}$, then for each input dataset~$x_{1:n} \in \cX^n$, we have 
    \begin{equation}
        e^{-\eps_0 / \paren{2n}} \le 
        \frac{
            \P{ \algoFiniteTime^{\cC} (x_{1:n}) \in Z }
        }{
            \P{ \algoPerfectPartitioning^{\cC} (x_{1:n}) \in Z }
        } 
        \le e^{\eps_0 / \paren{2n}}.
    \end{equation}
    Finally, for each pair of neighboring datasets~$x_{1:n}, x_{1:n}' \in \cX^n$ and each measurable~$Z$ in common output space~$\paren{\N^{\cY}}^n$, we have
    \begin{align*}
         \P{ \algoFiniteTime^{\cC} (x_{1:n}) \in Z }
            &\le e^{\eps_0 / \paren{2n}} \P{ \algoPerfectPartitioning^{\cC} (x_{1:n}) \in Z } \\
            &\le e^{\eps_0 / \paren{2n}} \left( e^\eps \P{ \algoPerfectPartitioning^{\cC} (x_{1:n}') \in Z } + \delta' \right)  \\
            &\le e^{\eps_0 / \paren{2n}} \left( e^\eps e^{\eps_0 / \paren{2n}} \P{ \algoFiniteTime^{\cC} (x_{1:n}') \in Z } + \delta' \right)  \\
            &= e^{ \eps + \eps_0 / n} \P{ \algoFiniteTime^{\cC} (x_{1:n}') \in Z } + e^{\eps_0 / \paren{2n}} \delta'\,. 
    \end{align*}
\end{proof}

\end{document}